\documentclass[aps,pre,twocolumn,amsmath,amssymb,showpacs,amsfonts]{revtex4}
\usepackage{epsfig}
\usepackage{graphicx}
\usepackage{dcolumn}
\usepackage{bm}
\usepackage{amsmath}
\graphicspath{{./Figures/}}

\newcommand{\bv}{{\bf v}}
\newcommand{\bg}{{\bf g}}

\newcommand{\bs}{\hat{\boldsymbol \sigma}}

\begin{document}
\title{Attempted density blowup in a freely cooling dilute granular gas: hydrodynamics versus molecular dynamics}
\author{Andrea Puglisi$^1$, Michael Assaf$^2$, Itzhak Fouxon$^2$, and Baruch Meerson$^2$}
\affiliation{$^1$Dipartimento di Fisica, Universit\`{a} La Sapienza, p.le Aldo Moro 2, 00185 Roma, Italy\\
$^2$Racah Institute of Physics, Hebrew University of Jerusalem,
Jerusalem 91904, Israel}
\begin{abstract}
It has been recently shown (Fouxon \textit{et al.} 2007) that, in the framework of ideal granular hydrodynamics (IGHD), an initially smooth hydrodynamic flow of a granular gas can produce an infinite gas density in a finite time.  Exact solutions that exhibit this property have been derived.  Close to the singularity, the granular gas pressure is finite and almost constant. This work reports molecular dynamics (MD) simulations of a freely cooling gas of nearly elastically colliding hard disks, aimed at identifying the ``attempted" density blowup regime. The initial conditions of the simulated flow mimic those of one particular solution of the IGHD equations that exhibits the density blowup. We measure the hydrodynamic fields in the MD simulations and compare them with predictions from the ideal theory. We find a remarkable quantitative agreement between the two over an extended time interval, proving the existence of the attempted blowup regime. As the attempted singularity is approached, the hydrodynamic fields, as observed in the MD simulations, deviate from the predictions of the ideal solution. To investigate the mechanism of breakdown of the ideal theory near the singularity, we extend the hydrodynamic theory by accounting separately for the gradient-dependent transport and for finite density corrections.

\end{abstract}

\maketitle

\section{Introduction}

Spontaneous clustering of particles in granular gases has attracted much recent interest \cite{Hopkins,Goldhirsch,McNamara1,McNamara2,Ernst,Brey,Luding,van
Noije,Ben-Naim2,ELM,MP,Garzo}. As other pattern-forming instabilities, the clustering
instability of a freely cooling granular gas has served as a sensitive probe of theoretical modeling and, first of all, of the Navier-Stokes granular hydrodynamics (GHD). Although the formal criteria of its validity may be quite restrictive, see below, GHD has a great power, sometimes far beyond its
formal validity limits \cite{Goldhirsch2}, in predicting a host of collective phenomena in granular flows, such as shocks, vortices and clusters.  In the recent years, GHD has been applied to a variety of \textit{non-stationary} dilute granular flows \cite{ELM,Bromberg,Volfson,Fouxon1,Fouxon2,MFV}. Non-stationary flows are both appealing and challenging for continuum modeling of granular dynamics. As in other areas of continuum modeling, this is especially true when a non-stationary flow develops a finite-time
singularitiy \cite{Kadanoff}. Examples are provided by the finite-time
blowup of the gas density: at zero gravity
\cite{ELM,Fouxon1,Fouxon2} (as described by the \textit{ideal} GHD discussed below, IGHD from now on), and at finite gravity \cite{Volfson} as described by the more complete, non-ideal GHD. Of course, a density blowup
in a gas with finite-size particles can only be an intermediate asymptotics, as the blowup is ultimately arrested: either by close-packing effects \cite{MP}, or by the gradient-dependent transport \cite{MFV}. Still, the attempted blowup regimes, signaling the development of high density
regions in the gas, are fascinating and worth a detailed investigation.  One such regime has been recently addressed by Fouxon \textit{et al.} \cite{Fouxon1,Fouxon2}. They dealt with a macroscopically one-dimensional, freely cooling, dilute granular flow in the framework of ideal GHD that neglects the gradient-dependent transport effects: the heat diffusion and viscosity. Fouxon \textit{et al.} derived a class of exact solutions to the ideal equations, for which an initially smooth flow develops a finite-time density blowup. Close to the blowup time $\tau$, the maximum gas density exhibits a power law behavior $\sim
(\tau-t)^{-2}$. The velocity gradient blows up as $\sim  - (\tau-t)^{-1}$, whereas
the velocity itself remains continuous and forms a cusp, rather than a shock
discontinuity, at the singularity. The gas temperature vanishes at the
singularity, but the pressure remains finite and almost constant. Extensive numerical simulations with the ideal hydrodynamic equations showed that the singularity, exhibited by the exact solutions, is universal, as it develops for quite general initial conditions \cite{Fouxon1,Fouxon2}.
The reason behind this universality is in the fact that the sound travel time through the region of the developing singularity is much shorter than the characteristic inelastic cooling time of the gas in that region. As a result, the pressure gradient (almost) vanishes in the singularity region, and the local features of this isobaric singularity become essentially independent of how the flow was initiated and how it behaves at large distances from the singularity. This singularity is of the same type as the one that develops, in the framework of the IGHD equations, in a general \textit{low Mach number} flow of a freely cooling granular gas \cite{MFV}.

Here we perform molecular dynamics (MD) simulations of a freely cooling gas of nearly elastically colliding hard disks, aimed at identifying the ``attempted" density blowup regime, predicted by the ideal analytical solutions \cite{Fouxon1,Fouxon2}. We simulate a freely evolving dilute gas of nearly elastically colliding hard disks in a narrow channel with perfectly elastic side walls. In this geometry both the clustering mode in the transverse directions, and the shear mode are suppressed (see Refs. \cite{Goldhirsch,McNamara1,ELM,MP} for detailed criteria).  As a result, the coarse-grained, or hydrodynamic flow depends only on the longitudinal coordinate along the channel (and time), as it was assumed in Refs. \cite{Fouxon1,Fouxon2}. We choose the initial conditions of the MD simulations so that the coarse-grained density, velocity and temperature fields are those producing one of the exact blowup solutions of the ideal GHD equations. Then we follow the time history of the hydrodynamic fields in the MD simulations and compare it with that predicted by the ideal exact solution and with numerical solutions of non-ideal GHD equations.

The remainder of this paper is organized as follows. In Section II we briefly summarize the  ideal GHD analysis \cite{Fouxon1,Fouxon2} of the density blowup: we present the ideal GHD equations, one of their exact solutions, its main features and expected limits of its validity. In Section III we describe our MD simulations and compare the hydrodynamic quantities, computed from the simulations, with the exact solution of the ideal GHD equations. We find that the exact solution is in remarkable quantitative agreement with the MD simulations over an extended time interval, proving the existence of the attempted density blowup regime. As the attempted singularity is approached, the hydrodynamic fields, as observed in the MD simulations, deviate from the predictions of the exact solution. To investigate the mechanism of breakdown of the ideal solution, we extend the hydrodynamic theory,  in Section IV, in two separate ways. In the first one we take into account the gradient-dependent transport: the heat diffusion and viscosity, but continue to assume that the gas is dilute. In the second one we neglect the gradient-dependent transport but take into account, in a semi-phenomenological way, finite density corrections. Section V summarizes our results and puts them into a more general context of hydrodynamic scenarios of clustering in freely evolving granular gases.

\section{Hydrodynamic theory and density blowup}

\subsection{Ideal granular hydrodynamics and exact solution}

We consider a two-dimensional granular gas of identical hard and smooth disks
with diameter $\sigma$ and mass set to unity, and adopt a simple model where the inelastic particle collisions are characterized by a constant coefficient of normal restitution $r$.
Throughout this paper we will only deal with nearly elastic collisions,
 \begin{equation}\label{firstineq}
     1-r \ll1\,,
 \end{equation}
and assume a very small Knudsen number:
\begin{equation}\label{secondineq}
l_{free}/L\ll1\,.
\end{equation}
Here $l_{free}$ is the mean free path of the particles, and $L$ is the characteristic length scale of the hydrodynamic fields that may depend on time. In addition, we will assume in this Section that the local gas density $\rho$ is much smaller than the close-packing density of disks $\rho_c=2/(\sqrt{3}\sigma^2)$:
\begin{equation}\label{thirdineq}
   \rho \sigma^2\ll 1\,.
\end{equation}
The strong inequalities (\ref{secondineq}) and (\ref{thirdineq}) need to be verified \textit{a posteriori}, once the hydrodynamic problem in question is solved.
The strong inequalities (\ref{firstineq})-(\ref{thirdineq}) enable one to employ the well-established
equations of Navier-Stokes granular hydrodynamics (see,\textit{ e.g.}
\cite{Goldhirsch2,BP}) that deal with three coarse-grained fields: the mass
density $\rho({\mathbf x}, t)$, the mean flow velocity ${\mathbf v}({\mathbf x},
t)$ and the granular temperature $T({\mathbf x}, t)$. In a sufficiently narrow channel
these fields depend only on the longitudinal coordinate $x$, and the hydrodynamic
equations become
\begin{eqnarray}&&
\frac{\partial \rho}{\partial t}+\frac{\partial(\rho v)}{\partial
x}=0, \label{hydrodynamics1}\\&&
\rho\left(\frac{\partial
v}{\partial t}+ v\frac{\partial v}{\partial
x}\right)=-\frac{\partial (\rho T)}{\partial
x}+\nu_0\frac{\partial}{\partial x}\left(\sqrt{T}\frac{\partial
v}{\partial x}\right), \label{hydrodynamics2}
\\&&
\frac{\partial T}{\partial t}+
v\frac{\partial T}{\partial x}=-T\frac{\partial v}{\partial
x}-\Lambda\rho T^{3/2}+\frac{\kappa_0}{\rho}\frac{\partial}{\partial
x} \left(\sqrt{T}\frac{\partial T}{\partial x}\right)\nonumber\\&&
+\frac{\nu_0\sqrt{T}}{\rho}\left(\frac{\partial v}{\partial
x}\right)^2\,,\label{hydrodynamics3}
\end{eqnarray}
where $\Lambda=\sqrt{\pi}(1-r^2) \sigma $,  $\nu_0=(2\sqrt{\pi}\sigma)^{-1}$ and
$\kappa_0=2/(\sqrt{\pi}\sigma)$, see \textit{e.g.} Refs. \cite{Brey,BP}.
Equations~(\ref{hydrodynamics1})-(\ref{hydrodynamics3})
differ from the
hydrodynamic equations for a gas of \textit{elastically} colliding disks only by
the presence in Eq.~(\ref{hydrodynamics3}) of the inelastic loss
rate term $-\Lambda \rho T^{3/2}$, that describes the
proportionality of the energy loss per particle to the number of
particle collisions per unit time (proportional to $\rho T^{1/2}$)
and to the energy loss per collision (proportional to $T$). This additional term, however,
brings a whole new physics (and mathematics) into the problem.

\begin{figure}
\includegraphics[width=8.5cm,clip=true]{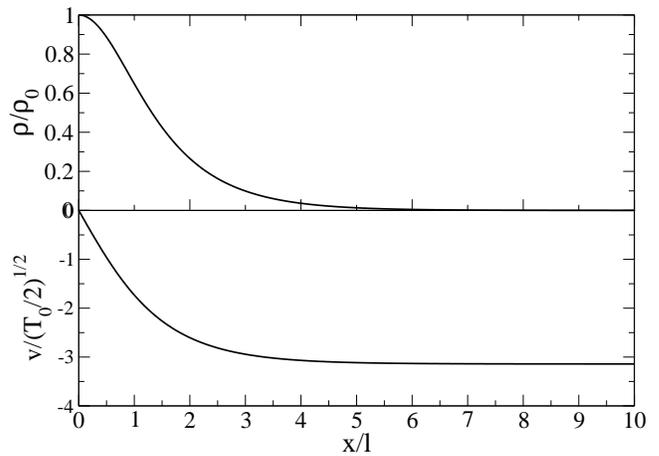}
\caption{
The initial values of the hydrodynamic fields in the example of attempted density blowup considered in this work. Shown are the rescaled initial density and velocity of the gas [see Eqs.~(\ref{incond1}) and (\ref{incond3})] versus the rescaled Eulerian coordinate $x/l$ along the channel. Only the right half of the system is shown. The values of $\rho_0$, $T_0$ and $l$, used in our molecular dynamic simulations, are presented in the beginning of subsection III.D.
\label{fig0}}
\end{figure}

Let us rewrite Eqs.~(\ref{hydrodynamics1})-(\ref{hydrodynamics3}) in
dimensionless variables. We will measure the gas density, temperature and the velocity in the units of $\rho_0$, $T_0/2$ and $\sqrt{T_0/2}$, respectively, where $\rho_0$ and $T_0$ are some characteristic values of the initial density and temperature. The time and distance will be measured in the units
of
\begin{equation}
\tau=\frac{4}{\Lambda\rho_0\sqrt{T_0}}\;\;\;\mbox{and}\;\;\;l=\tau\sqrt{\frac{T_0}{2}}\,,
\label{taul}
\end{equation}
respectively.  As one can see from Eq.~(\ref{hydrodynamics3}),  $\tau$ is the
characteristic cooling time of the gas due to the collisional energy loss, while $l$ is the characteristic distance a sound wave travels during time $\tau$. The numerical factors in Eqs.~(\ref{taul}) are chosen for convenience. We will keep identical notation for the rescaled and physical quantities, and take care that no confusion arises. Using the rescaled quantities, we rewrite Eqs.~(\ref{hydrodynamics1})-(\ref{hydrodynamics3}) as
\begin{eqnarray}&&
\frac{\partial \rho}{\partial t}+\frac{\partial
(\rho v)}{\partial x}=0,\label{hydro1}\\&&
\rho\frac{\partial v}{\partial t}+\rho
v \frac{\partial v}{\partial x} =-\frac{\partial
(\rho T)}{\partial x}+\frac{1-r^2}{4\sqrt{2}}
\frac{\partial}{\partial x}\left(\sqrt{T}\frac{\partial
v}{\partial x}\right),\label{hydro2}
\\&& \frac{\partial T}{\partial t}\!+\! v\frac{\partial T}
{\partial x}\!=\!- T\frac{\partial v}{\partial
x}\!-\!2\sqrt{2}\rho T^{3/2}\nonumber\\&&+\frac{1-r^2}{\sqrt{2}\rho}
\frac{\partial}{\partial x} \left(\sqrt{T}\frac{\partial
T}{\partial x}\right) +\frac{(1-r^2)\sqrt{T}}{4\sqrt{2}\rho}\left(\frac{\partial v}{\partial
x}\right)^2,\label{hydro3}
\end{eqnarray}
Let us assume that the characteristic magnitudes of the \textit{rescaled} hydrodynamic fields, and of their spatial and temporal derivatives,  are of order unity (this assumption needs to be checked \textit{a posteriori}). Then we can neglect the viscous and thermal conduction terms, as they scale as $1-r^2\ll 1$, and arrive at the IGHD equations:
\begin{eqnarray}&&
\frac{\partial \rho}{\partial t}+\frac{\partial(\rho v)}{\partial x}=0,\;\; \rho\left(\frac{\partial
v}{\partial t}+v\frac{\partial v}{\partial x}\right)= - \,\frac{\partial (\rho T)}{\partial x}, \label{a311} \\&& \frac{\partial T}{\partial t} + v\frac{\partial T}{\partial
x}=-T\frac{\partial v}{\partial
x}-2\sqrt{2} \rho T^{3/2}. \label{a322}
\end{eqnarray}
These equations were
investigated in Refs.\cite{Fouxon1,Fouxon2}, where a family of exact
analytic solutions was derived. Here we will consider a representative and simple
particular solution that evolves from the following
initial conditions:
\begin{eqnarray}&&
\rho(x, t=0)=\cosh^{-1} x,\ \ T(x,
t=0)=2,\label{initialconditions1}
\\&& v(x, t=0)=-2\arcsin
\left(\tanh x\right).\label{initialconditions2}
\end{eqnarray}
To remind the reader, we are using rescaled variables here. Back in the physical variables,
the initial profiles are
\begin{align}
\rho(x,0)&=\frac{\rho_0}{\cosh(x/l)},\label{incond1}\\
T(x,0)&=T_0,\label{incond2}\\
v(x,0)&=-\sqrt{2T_0} \arcsin\left[\tanh\left(\frac{x}{l}\right
)\right],\label{incond3}
\end{align}
where $l$ is defined in Eq.~(\ref{taul}). That is, at $t=0$ the
density profile has a maximum $\rho_0$ and width $l$, the temperature
$T_0$ is uniform, and there is an inflow of the gas towards the origin
with $v(x\to\pm\infty, t=0)=\mp\pi\sqrt{T_0/2}$. The initial scale of
variation of the fields, $l$, is by a factor $1/(1-r^2)$ greater than
the mean free path of the gas, justifying the use of the hydrodynamic
description. Furthermore, both the magnitudes, and the spatial scales
of the \textit{rescaled} fields are of order unity which justifies, at
least for finite times, the use of the ideal equations (\ref{a311})
and (\ref{a322}). Figure~\ref{fig0} shows the initial density and velocity fields of the flow in
the Eulerian coordinate.

Now we go over from the Eulerian coordinate $x$ to the Lagrangian mass
coordinate $m=\int_0^x \rho(x^{\prime}, 0) \,dx^{\prime}$, see \textit{e.g.} \cite{zeldovich}.
For the initial density profile (\ref{initialconditions1}), the Eulerian coordinate $x$ is related to the Lagrangian coordinate $m$  as follows:
\begin{equation}
x(m, t=0)=\frac{1}{2}\ln\left(\frac{1+\sin
m}{1-\sin m}\right). \label{fluidparticles2}
\end{equation}
Note that the infinite Eulerian interval $-\infty <x<+\infty$ corresponds, in this example, to a finite interval of $m$: $-\pi/2<m<\pi/2$, that is to a finite (rescaled) total mass of the gas, equal to $\pi$. In the Lagrangian coordinates, the ideal equations (\ref{a311}) and (\ref{a322}) are
\begin{eqnarray}&&
\frac{\partial}{\partial t} \left(\frac{1}{\rho}\right)=
\frac{\partial v}{\partial m},\,\,\, \frac{\partial v}{\partial t}
=-\frac{\partial p}{\partial m},\label{eqs1}\\&& \frac{\partial
p}{\partial t}=-2 p \rho\frac{\partial v}{\partial m}-2\sqrt{2}
p^{3/2}\rho^{1/2}\,, \label{eqs2}
\end{eqnarray}
where the dilute gas pressure $p=\rho T$ has been used instead of the temperature. The exact solution  in the Lagrangian coordinates is the following \cite{Fouxon1,Fouxon2}:
\begin{eqnarray}&&
\rho(m, t)\!=\!\frac{\cos m}{(1-t\cos m)^2},\ \ p(m, t)=2\cos m, \label{s2}\\&&  v(m, t)\!=\!-2 m+2t \sin
m. \label{s1}
\end{eqnarray}
As $x(m, t)$ can be calculated explicitly \cite{Fouxon2},
\begin{eqnarray}&&
x(m, t)=\frac{1}{2}\ln\left(\frac{1+\sin m}{1-\sin
m}\right)-2tm+t^2\sin m, \label{disp}
\end{eqnarray}
equations (\ref{s2})-(\ref{disp}) describe the time history of the hydrodynamic fields in the $x$-coordinate in a closed parametric form. Let us consider some important
features of this simple exact solution.

\subsection{Density blowup and its properties}

A momentary look at the first of Eqs.~(\ref{s2}) reveals that the density at the origin blows up in a finite time. Back in the dimensional variables one obtains
\begin{equation}
\rho(x=0, t)=\frac{\rho_0}{(1-t/\tau)^2}\,.\label{rhoorigin}
\end{equation}
The gas temperature becomes zero at the singularity, the velocity gradient $\partial v(x,t)/\partial x$ blows up as $\sim  - (\tau-t)^{-1}$, whereas
the velocity itself remains continuous and forms a cusp.   This behavior at singularity is quite different from that of the \textit{free flow} singularity (that one observes for the same initial velocity profile but a \textit{zero} gas pressure). In particular, for the free flow singularity the density blows up as $(1-t/\tau)^{-1}$, while the velocity develops a shock discontinuity \cite{Whitham}. See Refs. \cite{Fouxon1,Fouxon2} for more differences between the two types of the singularities.

It was found in Refs.~\cite{Fouxon1,Fouxon2} that the finite time blowup of the gas density and of the velocity gradient is not a consequence of specially chosen initial conditions. Rather, it appears, in the framework of the ideal equations (\ref{a311}) and (\ref{a322}) [or, equivalently, of
Eqs.~(\ref{eqs1}) and (\ref{eqs2})], for quite
general initial conditions. A robust local
feature of this singularity is a finite, non-zero value of the
gas pressure at the time of the density blowup. The singularity is universal because the sound travel time through it is much shorter than the characteristic inelastic cooling time of the gas. As a result, the pressure gradient (almost) vanishes in the singularity region, and the local features of this isobaric singularity become independent of the flow details at large distances.

As an infinite density cannot be reached in a gas with finite-size particles,  it is clear that,
sufficiently close to the attempted
singularity, some of the assumptions made on the way to the ideal equations
(\ref{a311})-(\ref{a322}) break down. However, independently
of the precise way of regularizing
the infinite density, an attempted singularity implies the formation of a region with a very high density. Furthermore, the ideal solution should accurately describe
the evolution of the system for times not too close to the attempted density blowup time $\tau$. Before we look into where the ideal theory breaks down, let us consider some
\textit{global} characteristics of the exact solution. For these the
singularity time $t=\tau$ turns out not to be special. First, we note that the
solution describes a gas with a (constant) finite number $N$ of
particles, given by
\begin{equation}
N=L_y\int_{-\infty}^{\infty} \rho(x) dx=\pi \rho_0 L_y l =
\frac{2\sqrt{2\pi}\, L_y}{(1-r^2)\sigma}\,,\label{numberparticles}
\end{equation}
where $L_y$ is the channel width. Note that $N$ is
independent of $\rho_0$: for larger $\rho_0$ the initial density profile
has a higher peak but a smaller width, so that $N$
remains constant. Another global characteristics of the solution is
the total energy of the gas
$$
E(t)=L_y\int_{-\infty}^{\infty} \left(\frac{\rho v^2}{2}+\rho T\right) dx\,.
$$
For the thermal part of the energy we find, after a simple algebra,
\begin{eqnarray}
\frac{E_{th}(t)}{L_y}&=& \int_{-\infty}^{\infty} \rho T dx \,= \,2 \int_{0}^{\infty} \rho T dx\nonumber \\
&=& 2\rho_0 T_0 l \int_0^{\pi/2} [1-(t/\tau)\cos m]^2 dm \nonumber \\
&=& \rho_0 T_0 l\left[\pi-4\left(\frac{t}{\tau}\right)+ \frac{\pi}{2}\left(\frac{t}{\tau}\right)^2\right].
\end{eqnarray}
Similarly, for the kinetic energy of the macroscopic motion we find
\begin{equation}
\frac{E_{kin}(t)}{L_y}=\rho_0 T_0
l\left[\frac{\pi^3}{12}-4\left(\frac{t}{\tau}\right)
+\frac{\pi}{2}\left(\frac{t}{\tau}\right)^2\right].
\end{equation}
Summing the two and using Eq.~(\ref{numberparticles}), we obtain \cite{Fouxon2}
\begin{eqnarray}&&
E(t)=N T_0
\left[1+\frac{\pi^2}{12}-\frac{8}{\pi}\left(\frac{t}{\tau}\right)
+\left(\frac{t}{\tau}\right)^2\right]\,. \label{energy}\end{eqnarray}
As time increases, $E(t)$ is monotone decreasing
for $t\leq \tau$. We also observe that $t=\tau$ is a regular point of $E(t)$, where nothing dramatic happens.  Note that the parabolic-in-time law of the energy decay is quite
different from Haff's law $E(t)=E(0)/(1+2t/\tau)^2$ obtained for freely cooling \textit{homogeneous} granular gas with density $\rho_0$
\cite{Haff}.

\subsection{Breakdown of ideal theory}

For a one-dimensional flow, the applicability of the solution (\ref{s2})-(\ref{disp}) is determined by the applicability of the ideal equations (\ref{a311})-(\ref{a322}) that it solves exactly. Analysis in Ref.~\cite{Fouxon2} shows that, sufficiently
close to the attempted singularity, the ideal equations become invalid. This happens because of one of two reasons (or both): either the gas becomes dense, so that criterion (\ref{thirdineq}) breaks down, or the heat diffusion becomes important, invalidating the ideal equations (\ref{a311})-(\ref{a322}). The time $t_{br}$ at which the ideal theory breaks down can be estimated as follows \cite{Fouxon2}:
\begin{equation}
1-\frac{t_{br}}{\tau}\sim \max(\sqrt{\rho_0\sigma^2}, 1-r).
\label{dominantmechanism}
\end{equation}
Therefore, the ``bottleneck" for the validity of the equations is set by the initial conditions: if the maximum is determined by the first (correspondingly, the second) term in the right hand side of Eq.~(\ref{dominantmechanism}),
the ideal equations become invalid because of the finite gas density (correspondingly, the finite heat diffusion). As each of the two terms is very small by assumption, the solution is expected to break down only close to the attempted singularity \cite{Knudsencheck}.

The one-dimensional solution may also become invalid because of instability with respect to small initial perturbations that are inevitably present in MD simulations.  Numerical solutions of the hydrodynamic equations, reported in Refs. \cite{Fouxon1,Fouxon2}, strongly suggest that the ideal solution is stable with respect to small \textit{longitudinal} perturbations. This does not exclude possible instability with respect to small \textit{transverse} perturbations. The only available analytic result here is the one obtained from the stability condition for a \textit{homogeneous} cooling state, see Refs. \cite{Goldhirsch,McNamara1,ELM,MP}. That stability criterion comes from a competition between the (destabilizing) inelastic cooling and the (stabilizing) heat diffusion and viscosity in the transverse direction. The stability criterion demands that $L_y$ be less than a threshold value depending on $1-r$, $\rho_0$ and $\sigma$. The stability problem for the strongly inhomogeneous and time-dependent exact solution is obviously more complicated, and its complete analytic solution
does not seem feasible. It is therefore important that our MD simulations, presented in the next Section, strongly suggest that, for sufficiently narrow channels, no instability in the transverse direction occurs for the time-dependent flow we are working with.

Let us summarize the main theoretical predictions. For the initial conditions (\ref{initialconditions1}) and (\ref{initialconditions2}), a nonlinear time-dependent flow sets in, described by the ideal exact solution: Eqs.~(\ref{s2})-(\ref{disp}). This flow ``attempts" to develop a density blowup. However, close to the attempted singularity one (or both) of the two factors: the finite density and the heat diffusion, invalidates the solution. The relative importance of the two factors is determined, via Eq.~(\ref{dominantmechanism}), by the initial conditions.

\section{Molecular Dynamics Simulations}

\subsection{General}

To test the theoretical predictions, we performed MD simulations of a
free cooling granular gas in a narrow two-dimensional
channel. The initial conditions correspond to hydrodynamic profiles
(\ref{initialconditions1}) and (\ref{initialconditions2}) and satisfy
the strong inequalities (\ref{firstineq})-(\ref{thirdineq}). According
to the theory, they are expected to
generate the nonlinear time-dependent flow described by
Eqs.~(\ref{s2})-(\ref{disp}). Our MD simulation calculates the
evolution of a gas of $N^{\prime}$ identical inelastic hard disks of
unit mass, with diameter $\sigma$, in a channel of dimensions
$L_x^{\prime}=L_x/2$ and $L_y$.  As the expected hydrodynamic flow is
symmetric with respect to $x=0$, only one half of the system, $x \in
[0,L_x/2]$, is simulated, so $N^{\prime}=N/2$.  Each wall of (the one
half of) the channel is solid and reflects elastically the disks
colliding with it. The particles move freely until a collision
(``event'') occurs when two disks $i$ and $j$ find themselves at a
distance equal to $\sigma$. The collision is resolved instantaneously,
leaving the positions of the particles unaltered and updating their
velocities from $(\bv_i,\bv_j)$, before the collision, to
$(\bv_i',\bv_j')$, after the collision. The update rule conserves the
total momentum and reduces the total kinetic energy, with a constant
coefficient of normal restitution $r \in [0,1]$:
\begin{align}
\bv_i'&=\bv_i-\frac{1+r}{2}(\bg \cdot \bs)\bs,\\
\bv_j'&=\bv_j+\frac{1+r}{2}(\bg \cdot \bs)\bs,
\end{align}
where $\bg=\bv_i-\bv_j$ and $\bs$ is the unit vector joining the
centers of the two disks. The hard-core interactions make possible
the following optimization of the algorithm. It is sufficient to
calculate the first collision times of all particles and then select
the absolute first one. The system is freely evolved up to that
time, then the collision is resolved, and a new list of collision
times is computed. With standard optimization techniques of the
search procedure it is possible to achieve fast computation times~\cite{poschel}.
Nevertheless, the time performance is proportional to the number of
collisions occurred, so the ratio between the physical time and
the cpu-time goes down when dense clusters emerge in the system.

\subsection{Initial conditions}

The initial position and velocity of each of the $N^{\prime}$ disks are chosen
randomly with probability distributions corresponding to the desired initial hydrodynamic fields.  This was implemented with the following procedure. For each disk $i$

\begin{enumerate}

\item
the longitudinal position $x_i$ is chosen with probability
proportional to $\rho(x,0)$ from Eq.~(\ref{incond1}) for $x\geq0$, using an acceptance/rejection method:

\begin{enumerate}

\item \label{x_extr} a random $x_i$-position is generated with uniform
probability on the interval $[\sigma/2,L_x^{\prime}-\sigma/2]$,

\item
a random number $z$ with uniform probability on
$[0,\max\{\rho(x,0)\}]$ is compared with $\rho(x_i,0)$,

\item
if $z<\rho(x_i,0)$ the position is accepted, otherwise the
procedure is repeated from~\ref{x_extr},

\end{enumerate}

\item
then the vertical position $y_i$ is chosen with uniform probability
on the interval $[\sigma/2,L_y-\sigma/2]$,

\item
a non-overlap check is performed: the distance between the disk
center $(x_i,y_i)$ and all the previously placed disk centers must
be greater then $\sigma$: if the condition is not satisfied, the
procedure is repeated from~\ref{x_extr},

\item
the velocity components $v^x_i$ and $v^y_i$ are chosen from a Gaussian
distribution with zero mean and variance equal to $T_0$; then the
longitudinal component is shifted by an amount $v(x,0)$ from Eq.~(\ref{incond3}) with $x\geq0$.

\end{enumerate}

\subsection{Lagrangian coordinate and hydrodynamic fields}

We verified that, for our choice of the channel dimensions, the gas remained
homogeneous in the $y$-direction. By virtue of this observation,
it was sufficient to deal with
one-dimensional hydrodynamic profiles depending on $x$.
For a direct comparison with the
analytical solution of the IGHD, the hydrodynamic profiles were obtained using
a uniform binning in the Lagrangian mass coordinate. Using the same notation as in Section I,
we define the Lagrangian mass interval for the simulated flow
as $[0,\pi/2$], where $\pi/2$
corresponds to (one half of) the total gas mass, $N^{\prime}=N/2$.
Let $n_{bin}$ be the number of bins chosen to sample the
hydrodynamic profiles and $N_{bin}=N^{\prime}/n_{bin}$ be the average number of
particles per bin. At a given time $t$ all particles are ordered so
that $x_i<x_{i+1}$, $i \in [0,N^{\prime}-1]$. Then each bin $j \in
[1,n_{bin}]$ has its leftmost border at $x_{[(j-1)N_{bin}]}$ and
its rightmost border at $x_{[jN_{bin}-1]}$. These bins are
non-uniform in the $x$ coordinate, but are uniform in the
Lagrangian mass coordinate as each contains the same mass
$N_{bin}$. The position of the $j$-th
bin is
$$
m_j=\frac{\pi (j-1/2)}{2 n_{bin}}.
$$
All particles
belonging to the $j$-th bin contribute to the value of the
hydrodynamic fields:
\begin{eqnarray}
  &&\rho(m_j,t) = \frac{N_{bin}}{L_y L_j}, \\
  &&v(m_j,t) = \frac{\sum_{i \in j} v^x_i}{N_{bin}}, \\
  &&T(m_j,t)= \frac{\sum_{i \in j} \left[(v^x_i)^2+(v^y_i)^2\right]}{2 N_{bin}}-v^2(m_j,t),
\end{eqnarray}
where we have used the shorthand notation $i \in j$ to denote
particles in the $j$-th bin, and $L_j$ to denote the length of the
$j$-th bin. The pressure field $p(m_j,t)=\rho(m_j,t) T(m_j,t)$ is
obtained straightforwardly.

The hydrodynamic fields, computed for individual realizations, exhibit
a significant noise. To get rid of the noise, all hydrodynamic profiles presented in the next Section
were obtained after an averaging over $100$ MD
simulations with different initial conditions, corresponding to the same initial hydrodynamic fields and obtained with the procedure described in subsection B.

\subsection{MD simulations versus ideal solution}

\begin{figure}
\includegraphics[width=8.5cm,clip=true]{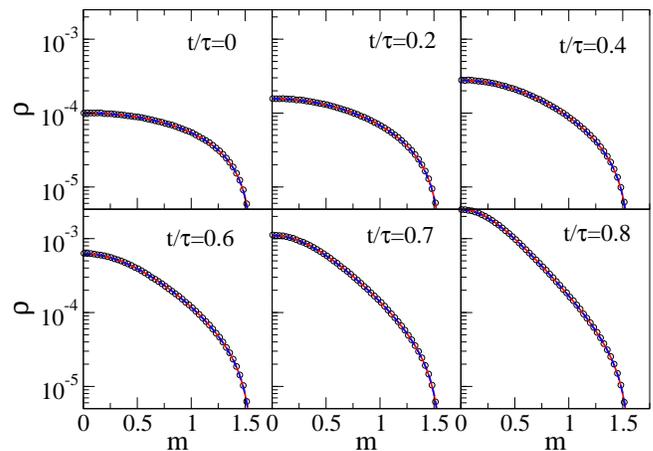}
\caption{(Color online) Evolution of the density $\rho(m,t)$ in the Lagrangian frame
at initial times. The circles: the results of MD simulations. The solid line: the
prediction of the analytic solution, first of Eqs.~(\ref{s2}), back in the physical variables. The dashed line:
a numerical solution of the non-ideal hydrodynamic equations (\ref{hydro1})-(\ref{hydro3}) that account for the gradient-dependent transport. The dashed line is indistinguishable from the solid line. \label{fig:rho_0_700}}
\end{figure}

\begin{figure*}
\includegraphics[width=11cm,clip=true]{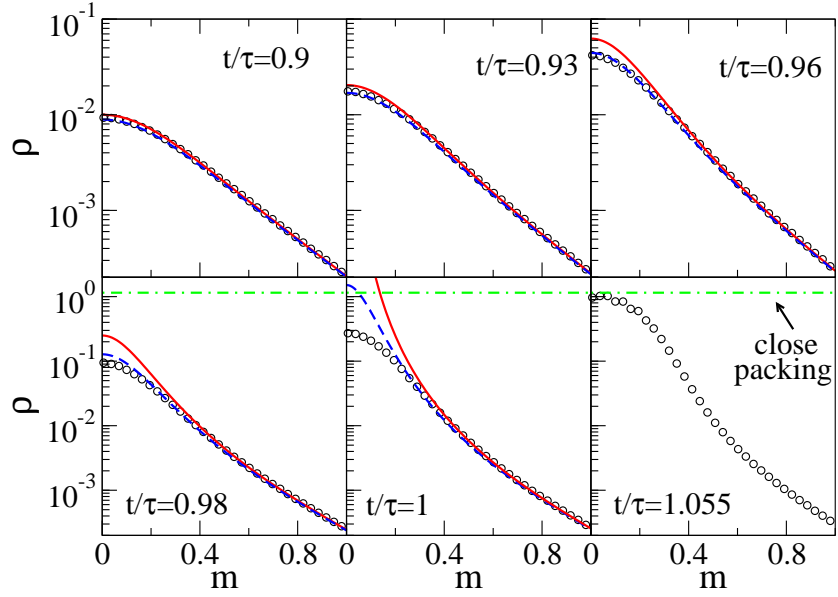}
\caption{(Color online) Evolution of the density $\rho(m,t)$ in the Lagrangian frame
at late times. The $m$-axis is zoomed in, in order
to focus on the high density region. The circles: the results of MD simulations.
The solid line: the prediction of the first of Eqs.~(\ref{s2}), back in the physical variables. The
dashed line: a numerical solution of the non-ideal hydrodynamic equations (\ref{hydro1})-(\ref{hydro3}) that account for the gradient-dependent transport.  The
dash-dotted line marks the close packing density
$\rho_c=2/(\sqrt{3}\sigma^2)$.\label{fig:rho_800_900}}
\end{figure*}

\begin{figure}
\includegraphics[width=8.5cm,clip=true]{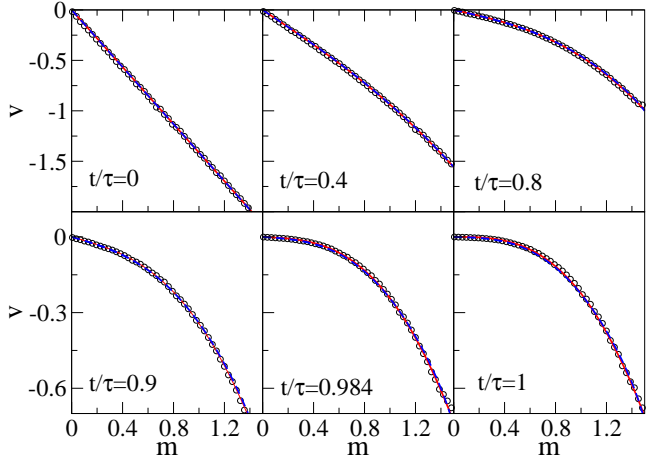}
\caption{(Color online) Evolution of the longitudinal velocity $v(m,t)$ in the
Lagrangian frame.  The circles: the results of MD simulations. The solid line:
the prediction of Eq.~(\ref{s1}), back in the physical variables. The
dashed line: a numerical solution of the non-ideal hydrodynamic equations (\ref{hydro1})-(\ref{hydro3}) that account for the gradient-dependent transport. The dashed line is indistinguishable from the solid line. \label{fig:vx_0_900}}
\end{figure}

\begin{figure}
\includegraphics[width=8.5cm,clip=true]{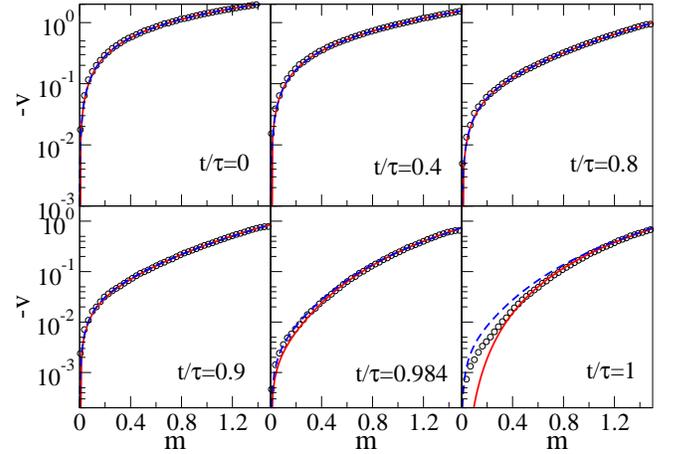}
\caption{(Color online) Evolution of the (minus) longitudinal velocity $v(m,t)$ in the
Lagrangian frame, in a semi-logarithmic scale. The circles: the results of MD simulations.
The
solid line: the prediction of
Eq.~(\ref{s1}), back in the physical variables. The dashed line: a numerical
solution of the non-ideal hydrodynamic equations (\ref{hydro1})-(\ref{hydro3}) that account for the gradient-dependent transport.\label{fig:vx_log_0_900}}
\end{figure}

The following parameters were chosen for the simulations: $\sigma=1$, $\rho_0=10^{-4}$, $T_0=1$, $N^{\prime}=5\times10^4$, and $L_y=125$. For convenience, the coefficient of normal restitution $r$ was chosen so that $1-r^2=\sqrt{\pi/2}\times 10^{-2}$, i.e. $r = 0.99371367\dots$. This choice of parameters sets $l \simeq 1.27324 \times 10^6$, $L_x^{\prime}=10 l$ and $\tau \simeq 1.8006 \times 10^6$. The evolution of the density field, as obtained in the simulations and as predicted by the ideal theory, is displayed in
Fig.~\ref{fig:rho_0_700} for times up to $t=0.8 \tau$, and in
Fig.~\ref{fig:rho_800_900} for later times, up to
time $t=1.055\tau$.
The figures show that
the ideal solution is in remarkable agreement with the MD simulations up to times $t\simeq
0.9 \tau$. At later times, when the density peak exceeds $\simeq 10^{-2}$, the ideal solution starts to deviate from the MD simulation
in the
neighborhood of $m=0$. The actual density peak continues to grow, but
slower than predicted by the ideal solution. At time $t=\tau$, when the ideal solution predicts the density blowup in $x=0$, the actual density $\rho(0,\tau) \simeq
0.2$. The close-packing density $\rho_c=2/(\sqrt{3}\sigma^2)$ is reached at
$t \simeq 1.05 \tau$, see the last frame of
Figure~\ref{fig:rho_800_900}. Sufficiently far from $m=0$, the ideal solution
remains very accurate.  (We checked that this statement remains true even beyond the attempted singularity time:  until the end of the MD simulations.)

\begin{figure}
\includegraphics[width=8.5cm,clip=true]{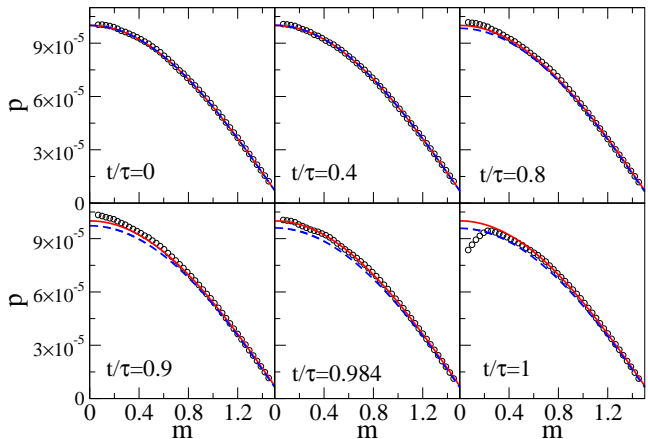}
\caption{(Color online) Evolution of the gas pressure, obtained
using the ideal equation of state $p(m,t)=\rho(m,t)T(m,t)$,
in the Lagrangian frame.  The circles: the results of MD simulations.
The solid line: the prediction of the second of Eqs.~(\ref{s2}), back in the physical variables. The
dashed line: a numerical solution of the non-ideal hydrodynamic equations (\ref{hydro1})-(\ref{hydro3}) that account for the gradient-dependent transport.
\label{fig:pr_0_900}}
\end{figure}

The gas velocity profiles, shown in Figs.~\ref{fig:vx_0_900}
and~\ref{fig:vx_log_0_900} in a linear and logarithmic scale, respectively, are very accurately predicted
by the ideal solution, Eq.~(\ref{s2}), until late times.
Surprisingly, the excellent agreement remains even
at times greater than $0.9 \tau$, when the density peak already significantly
deviates from the theoretical one. To be able to see the small deviations from the theory, we had to use, in Fig.~\ref{fig:vx_log_0_900}, a logarithmic scale.

Similarly, an inspection of the pressure
profiles, see Fig.~\ref{fig:pr_0_900}, shows an excellent
agreement with the prediction of the ideal theory, $p(m,t)=\rho_0 T_0 \cos(m)$,
see the second of Eqs.~(\ref{s1}). Discrepancies
of about $2\%$ appear only at late times, when the density
is already about one half of the theoretically predicted value. At times close to $\tau$, the
pressure field, as found in the MD simulations, develops a dip close to $x=m=0$, reaching a
value about $15\%$ lower than the theoretically expected value $p(0)=10^{-4}$.

\begin{figure}
\includegraphics[width=7.0cm,clip=true]{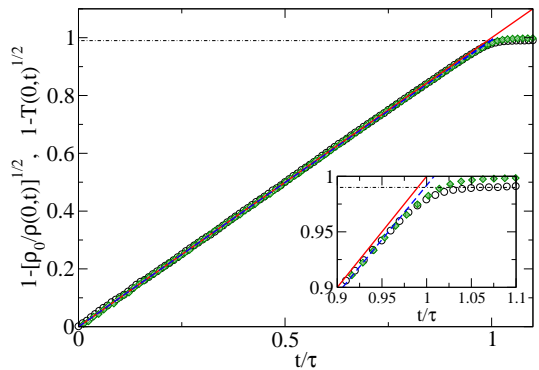}
\caption{(Color online) The time history of the density maximum $\rho(0,t)$ and
the temperature minimum $T(0,t)$. The symbols show the results of MD simulations: black circles depict the quantity $1-\sqrt{\rho_0/\rho(0,t)}$, green diamonds  depict the quantity $1-\sqrt{T(0,t)}$. The solid line is the prediction of Eq.~(\ref{rhoorigin1}): an immediate consequence of Eq.~(\ref{rhoorigin}).
The dashed line is a numerical solution of the non-ideal hydrodynamic equations (\ref{hydro1})-(\ref{hydro3}) that account for the gradient-dependent transport. The inset zooms in at later times. \label{fig:rhomax}}
\end{figure}

A direct characterization of the attempted gas density blowup is provided by the time history of the density at $x=0$. The ideal solution predicts, see Eq.~(\ref{rhoorigin}), that
\begin{equation}
1-\sqrt{\frac{\rho_0}{\rho(0,t)}}=\frac{t}{\tau}.
\label{rhoorigin1}
\end{equation}
This prediction is extremely well supported by the MD simulations in Fig.~\ref{fig:rhomax}, until  $t\simeq 0.9 \tau$. The subsequent deviation from the theory appears
as saturation of the quantity $1-\sqrt{\rho_0/\rho(0,t)}$ at the
value $1-\sqrt{\rho_0/\rho_c}$ corresponding to the close packing density $\rho_c$.
The same Fig.~\ref{fig:rhomax} also depicts a different quantity,
$1-\sqrt{T(0,t)}$. In view of the theoretical
expectation $p(0,t)=\rho_0 T_0 =\rho_0$, this quantity is also expected to grow
as $t/\tau$, and Fig.~\ref{fig:rhomax} indeed shows this growth until $t\simeq 0.9 \tau$. The quantity $1-\sqrt{T(0,t)}$ also
saturates at later times, but at a value slightly different from $1-\sqrt{\rho_0/\rho_c}$, see the inset  of Fig.~\ref{fig:rhomax}. This is consistent with the pressure deviation from $\rho_0$ at very late times.

\begin{figure}
\includegraphics[width=7.0cm,clip=true]{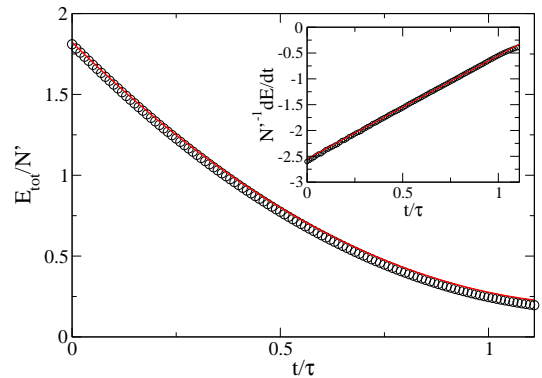}
\caption{(Color online) Decay of the total energy per particle $E(t)/N^{\prime}$ as measured in the MD simulations (the
circles) and predicted by the exact solution (the solid line).
The time derivative $(N^{\prime})^{-1} dE/dt$, displayed in the inset (the notation for the circles and line is the same), shows that the energy decay remains smooth at all simulated times. \label{fig:ene}}
\end{figure}

We also present, in
Fig.~\ref{fig:ene}, the time history of the total energy per particle,
\begin{equation}
\frac{E(t)}{N^{\prime}}=\frac{1}{N^{\prime}}\sum_{i=1}^{N^{\prime}} \frac{|v|^2}{2},
\end{equation}
as found in the MD simulations, and compare it with the theoretical prediction, Eq.~(\ref{energy}). Here the agreement is very good at all times, with a $3\%$ error at late times. The (numerical) time derivative of $E(t)$, depicted in the inset of Fig.~\ref{fig:ene}, remains smooth also close to the singularity time $\tau$. Actually, this is not surprising, as the main contribution to the thermal energy of the gas is made by the peripheral gas (in the Lagrangian frame), which is hotter and more dilute than the gas in the region close to $m=0$. Furthermore, the main contribution to the kinetic energy of macroscopic motion is again made by the peripheral gas (in the Lagrangian frame) which moves faster than the gas in the region close to $m=0$. The peripheral gas continues to follow the ideal theory at \textit{all} simulation times, and this explains the remarkable success of the ideal solution in predicting the total energy history.

To conclude this Section, the MD simulations clearly show,  over an extended period of time, the existence of the ``attempted" density blowup regime. The ideal granular hydrodynamics (IGHD) predicts very accurately the
hydrodynamic profiles, observed in the MD simulations, up to times close to the attempted singularity. The density field,
as measured in the MD simulations,
starts to deviate from the ideal theory at time  $t \simeq 0.9 \tau$. Somewhat surprisingly,  the rest of the hydrodynamic fields continue to show good agreement with the theory until even closer to the attempted singularity time $\tau$. In the following
Section we will see that the agreement with theory at later times improves significantly  when
the non-ideal hydrodynamic equations (\ref{hydro1})-(\ref{hydro3}), that account for the gradient-dependent transport, are employed.

\section{Non-ideal hydrodynamics}

To investigate the mechanism of breakdown of the ideal theory, we extended the hydrodynamic theory in two separate ways. In the first one we took into account the gradient-dependent transport: the heat diffusion and viscosity, but continued to assume the the gas is dilute. In the second one we neglected the gradient-dependent transport but took into account finite density corrections. The hydrodynamic equations were solved numerically in Lagrangian coordinates, using an accurate variable
mesh and variable time step solver \cite{blom}.

First, we solved (the Lagrangian form of) Eqs.~(\ref{hydro1})-(\ref{hydro3}) of non-ideal granular hydrodynamics (NIGHD). These equations account for the viscous and heat diffusion terms but still assume a dilute gas.
\begin{figure}
\includegraphics[width=7.0cm,clip=true]{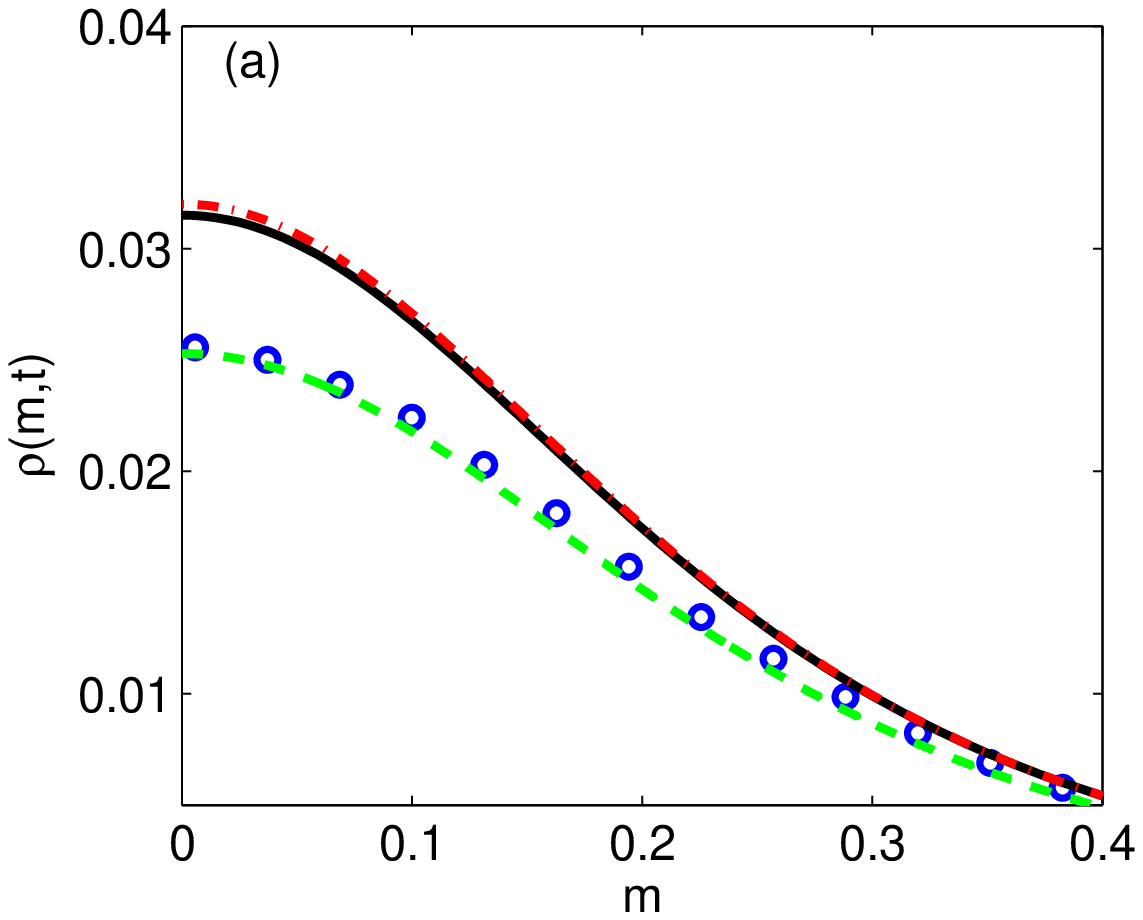}
\includegraphics[width=7.0cm,clip=true]{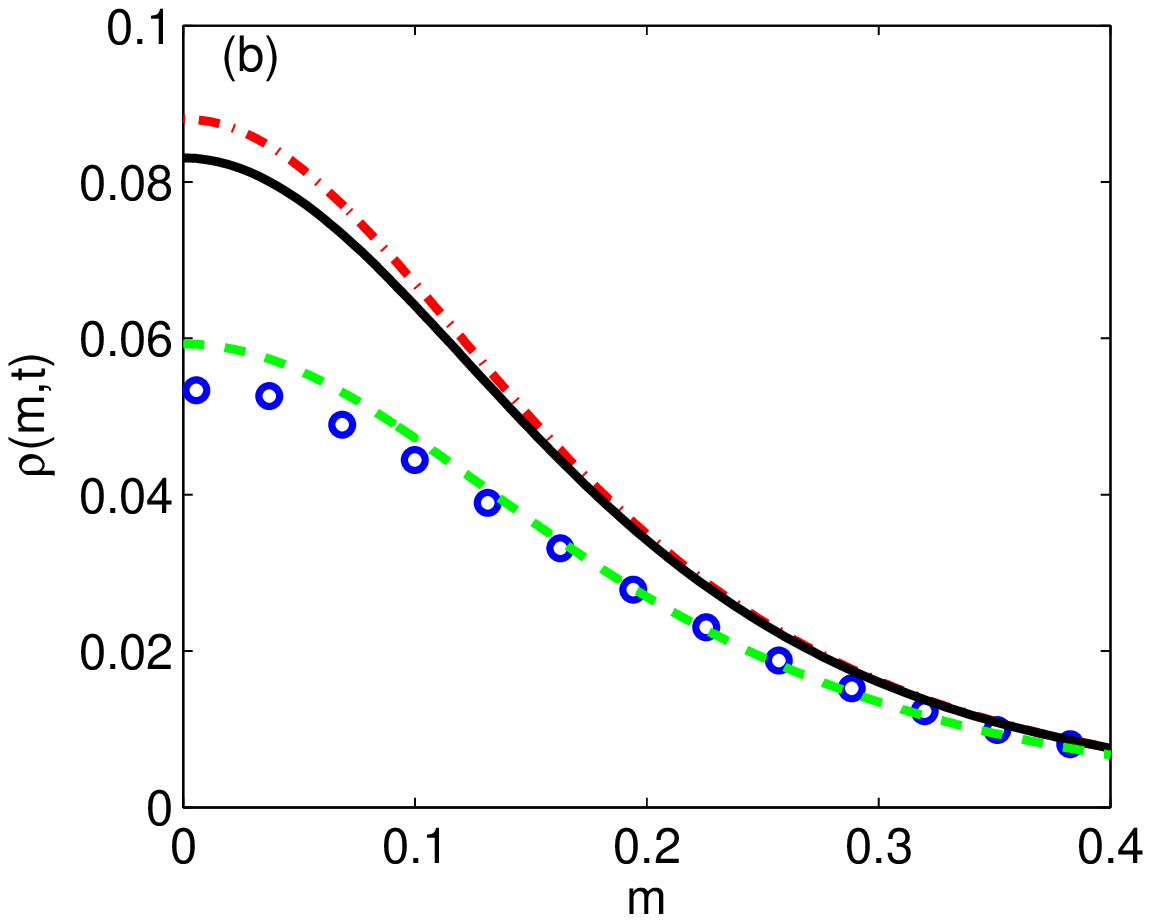}
\caption{(Color online) The density profiles for $t/\tau=0.944$ (a) and
$t/\tau=0.966$ (b) as obtained from the MD simulations (the circles), the analytic solution (the dashed-dotted line), the
numerical solution of the dilute NIGHD equations (the dashed line), and the numerical solution of finite-density hydrodynamic equations with the gradient-dependent transport terms neglected (the solid line). }\label{enskvisc}
\end{figure}

The numerically obtained NIGHD profiles are presented, together with the MD
simulations and the ideal analytical solution, in Figs.
\ref{fig:rho_0_700}-\ref{fig:rhomax}. As expected, the NIGHD profiles
coincide with the MD simulations and with the ideal theory for early and
intermediate times. At late times, the gradient-dependent transport terms become significant, and the NIGHD profiles approximate the MD simulation results much better than the ideal theory. As the maximum density continues to increase (and finally approaches the close packing density $\rho_c$), the dilute NIGHD description ultimately breaks down. In Fig. \ref{fig:rho_800_900} it occurs at about $t/\tau\simeq 0.98$.

In the second type of hydrodynamic computations we discarded the viscous and heat diffusion terms but took into account (moderate) finite density effects. This was done by adopting, in Eqs.~(\ref{eqs1}) and (\ref{eqs2}), instead of the ideal equation of state and ideal energy loss rate,  the Carnahan-Starling equation of state \cite{Carnahan} and a modification of the energy loss rate, derived by Jenkins and Richman \cite{jenkins} in the spirit of Enskog theory:
\begin{eqnarray}
p&\to&\rho T\left[1+\frac{\pi\rho}{\sqrt{3}}\,g(\rho)\right]\,,\nonumber\\
\Lambda&\to&\Lambda g(\rho)\,,
\end{eqnarray}
where
$$
g(\rho)=\frac{1-\frac{7\pi\rho}{32\sqrt{3}}}{\left(1-\frac{\pi\rho}{2\sqrt{3}}\right)^{2}}
$$
is the equilibrium pair correlation function of hard disks at contact. In the dilute limit $\rho \to 0$ one obtains $g=1$ and recovers the ideal equation of state $p=\rho T$.

In Fig.~\ref{enskvisc} we compare four different results for the gas density: the MD simulations, the ideal analytical solution, the numerical solution of the first type (the NIGHD-equations), and the numerical solution of the second type. It is clearly seen that, for the choice of parameters used in our MD simulations, the numerical solution of the second type is not as successful as that of the first type.  This could be expected, as for the times when the maximum density
is still much smaller than the close packing density, the
finite-density corrections (which are of the order of $\rho/\rho_c$) are still negligible. In contrast, the numerical results from the NIGHD equations agree well with the MD simulations and show that, as the attempted density blowup is approached, the heat conduction and viscosity effects can become important when the gas density is still small.

\section{Summary and Discussion}

Our MD simulations proved the existence of an attempted density blowup
regime as described by an exact solution of the ideal granular
hydrodynamic equations. We found the ideal solution to be in
remarkable quantitative agreement with the MD simulations over an
extended time interval, but not too close to the attempted
singularity. As the attempted singularity is approached, the exact
solution breaks down. A more complete hydrodynamic theory,
that accounts for the heat diffusion and viscosity, but still assumes
a dilute gas, continues to agree with the MD simulations
until the gas density becomes a fraction of the close packing density
of disks.

Let us put the results of this work into a more general context of
clustering instability of a freely cooling granular flow. As we have
already noted, the local properties of the density blow-up, exhibited by the exact
solutions of the IGHD equations, are the same as the local properties
of the density blow-up exhibited by a \textit{low Mach number} flow of a freely cooling
granular gas with the heat diffusion neglected \cite{MFV}. A low Mach number flow emerges when the
pressure balance sets in on a shorter time scale than the temperature
balance. In this case any local inelastic cooling causes a (low Mach
number) gas inflow into the colder region so as to increase the local
gas density there and keep the pressure gradient (almost) zero. The
resulting density instability develops on the background of an
(almost) homogeneous gas pressure. This is consistent with the MD
simulations presented here, see Fig.~\ref{fig:pr_0_900}, where the pressure in the
vicinity of the density maximum hardly changes up to times very close to the
attempted singularity time. For brevity we will call the low Mach number flow
instability scenario Scenario 1. Scenario 1 first appeared in
astrophysics and plasma physics in the context of condensation
instabilities in gases and plasmas that cool by their own radiation
\cite{Meerson}.

As many as \textit{four} additional hydrodynamic scenarios of clustering in a freely cooling granular gas have been discussed in the literature. We start with the \textit{pressure instability} scenario, or Scenario 2. It was discussed, in the context of the granular clustering, by  Goldhirsch and Zanetti \cite{Goldhirsch}, although it was also known earlier to the astrophysics and plasma physics communities, see Ref. \cite{Meerson} for a review. Scenario 2 is usually presented in the following way. Let us consider a small local increase in the gas density. This increase causes an increase in the collisional energy loss. As a result, the gas pressure falls down, a gas inflow develops, causing a further density increase, and the process continues. Importantly, Scenario 2 assumes that the inelastic cooling time is much shorter than the sound travel time. In other words, it is the local temperature balance that sets in rapidly here, and the resulting pressure gradient drives the flow on a relatively slow time scale.

As of present, there has been no detailed nonlinear analysis behind Scenario 2. The (well established) linear stability theory of the homogeneous cooling state \cite{McNamara1,Ernst} indicates that Scenarios 1 and 2 operate in two opposite limits: for sufficiently short and long perturbation wavelengths, respectively. The physics behind this is the following. The characteristic cooling time due to the inelastic collisions is independent of the length scale of the initial perturbation, whereas the sound travel time scale is proportional to it. As a result, when all other parameters are fixed, Scenario 1 corresponds to an intermediate wavelength limit of the clustering instability, while Scenario 2 corresponds to the long wavelength limit. (In the short wavelength limit the homogeneous cooling state of the gas is stable, as the clustering instability is suppressed by the heat diffusion \cite{McNamara1,MFV}.)

Now let us consider Scenario 3 that also assumes a long-wavelength limit. As the gas temperature falls down rapidly because of the inelastic cooling, the flow is describable by the zero pressure (or flow by inertia) approximation \cite{ELM}. Were it not regularized by close packing effects, such a flow would develop a finite-time density blowup (of a different type than the low Mach number flow) \cite{ELM,Whitham}. If the compressional heating interferes earlier than the close packing effects, the pressure becomes relevant again, and Scenario 3 gives way to the Scenario 1 \cite{Fouxon1}. In the opposite case the late time dynamics of the system is describable by the Burgers equation \cite{MP}. Which of the two regimes is realized in a particular setting depends on the initial conditions.

Scenarios 1-3 do not invoke the shear mode instability, and so they can operate both in one-dimensional, and multi-dimensional settings. On the contrary, Scenarios 4 and 5 do invoke the shear mode, and so they are intrinsically multi-dimensional (and intrinsically non-linear). Scenario 4 exploits the fact that the unstable shear mode may contribute, via a non-linear coupling, to the growth of the clustering mode. Obviously, the nonlinear coupling is the \textit{only} hydrodynamic mechanism of driving the clustering mode if the system size is larger than the critical size for the shear mode instability, but smaller than the critical size for the clustering mode instability. Furthermore, numerical analysis, performed in Ref. \cite{Brey}, indicated  that the nonlinear coupling plays a dominant role in the initial density growth also in the case  when the system size is \textit{comparable} to the critical system sizes for the clustering and shear instabilities. (More precisely, the wave number of the monochromatic test perturbation of the transverse velocity in Ref. \cite{Brey} was within the instability regions of both the shear, and the clustering modes. However, \textit{twice} the wave number already came out of the instability region.) What happens in much larger systems is presently under investigation. It turns out that \textit{well} above the clustering mode instability threshold the nonlinear coupling with the shear mode does \textit{not} dominate the density growth (though it does make the theory more cumbersome). In the channel geometry, that we adopted in this paper and in the previous works \cite{ELM,MP,Fouxon1,Fouxon2,MFV}, the shear mode is suppressed, and the clustering instability develops in its pure and simplest form.

Now we proceed to Scenario 5 \cite{Goldhirsch} that exploits the fact that the unstable shear mode \textit{heats} the gas in some regions.  Scenario 5 assumes  that this heating can be balanced by the inelastic energy loss, rendering (quite a complicated) steady state. It is furthermore assumed that this steady state is unstable with respect to small perturbations, and it is this instability that causes the granular clustering. We are unaware of a quantitative theory that would support Scenario 5, or of any quantitative test of Scenario 5 in MD simulations or in numerical solutions of hydrodynamic equations.

To complete the comparison of the five hydrodynamic scenarios of clustering we note that the only scenarios that have addressed, up to date,  a strongly nonlinear stage of
the clustering process \textit{quantitatively} is the low Mach number flow instability scenario (Scenario 1) \cite{MFV} and the zero pressure scenario (Scenario 3) \cite{ELM,MP}. It is the consideration of a strongly nonlinear stage that enables one to identify attempted finite-time density blowups: prototypes of the dense granular clusters.

Which results of this work will withstand a generalization to more realistic granular flow conditions: for example, rotational degrees of freedom and tangential inelasticity
of collisions?  Including the rotational degrees of freedom and tangential inelasticity
of collisions into a hydrodynamic description is possible under some
limitations \cite{BP,Goldhirsch2}. Solving the corresponding nonlinear hydrodynamic equations
analytically will of course be beyond our reach. It is likely that,
when the gradient-dependent transport is negligible, these nonlinear
equations will again exhibit a finite-time density blowup. Indeed,
the development of closely packed granular clusters in a granular flow is
a robust phenomenon. Therefore, it is natural to conjecture,
based on results of this work, that more realistic granular
clusters (those emerging when the rotational degrees of freedom are taken
into account) will still be describable as regularized attempted density
blowups.

In summary, the results of the present work gives support to the notion of a granular cluster
as a regularized density blowup of ideal granular hydrodynamic equations, put forward in Refs.
\cite{ELM,MP,Volfson,Fouxon1,Fouxon2,MFV}. In more general terms, they present additional evidence
that granular hydrodynamics is a powerful and
accurate quantitative theory of granular flows, especially once it is employed within its limits of applicability but, luckily, sometimes even beyond them.

\begin{acknowledgments}
Our work was supported by the Italian MIUR (PRIN grant
n. 2005027808$\_$003), by the Israel Science Foundation (grant
No. 107/05) and by the German-Israel Foundation for Scientific
Research and Development (Grant I-795-166.10/2003).
\end{acknowledgments}


\begin{thebibliography} {99}

\bibitem{Hopkins} M. A. Hopkins and M. Y. Louge, Phys. Fluids A \textbf{3}, 47 (1991).
\bibitem{Goldhirsch} I. Goldhirsch and G. Zanetti, Phys. Rev. Lett. \textbf{70}, 1619 (1993); I. Goldhirsch, M.-L. Tan, and
G. Zanetti,  J. Sci. Comp. \textbf{8}, 1 (1993).
\bibitem{McNamara1} S. McNamara, Phys. Fluids A \textbf{5}, 3056 (1993).
\bibitem{McNamara2} S. McNamara and W. R. Young,  Phys. Rev. E \textbf{53}, 5089 (1996).
\bibitem{Ernst} R. Brito and M. H. Ernst,  Europhys.
Lett. \textbf{43}, 497 (1998).
\bibitem{Brey}  J. J. Brey,  M. J. Ruiz-Montero, and D. Cubero, Phys. Rev. E \textbf{60}, 3150 (1999).
\bibitem{Luding} S. Luding and H. J. Herrmann,  Chaos \textbf{9}, 673 (1999).
\bibitem{van Noije} T.P.C. van Noije and M.H. Ernst, Phys. Rev. E \textbf{61}, 1765 (2000).
\bibitem{Ben-Naim2}  X. B. Nie, E. Ben-Naim, and S. Y. Chen,  Phys. Rev. Lett. \textbf{89}, 204301 (2002).
\bibitem{ELM}  E. Efrati, E. Livne, and B. Meerson,  Phys. Rev. Lett. \textbf{94},
088001 (2005).
\bibitem{MP} B. Meerson and A. Puglisi, Europhys. Lett. \textbf{70}, 478 (2005).
\bibitem{Garzo} V. Garz\'{o}, Phys. Rev. E \textbf{72}, 021106 (2005).
\bibitem{Goldhirsch2} I. Goldhirsch,  Annu. Rev. Fluid Mech. \textbf{35},
267 (2003).
\bibitem{Bromberg} Y. Bromberg, E. Livne, and B. Meerson,  in \textit{Granular Gas
Dynamics}, edited by T. P\"{o}schel and N.V. Brilliantov (Springer, Berlin,
2003), p. 251; cond-mat/0305557.
\bibitem{Volfson} D. Volfson, B. Meerson, and L. S. Tsimring,  Phys. Rev. E  \textbf{73}, 061305 (2006).
\bibitem{Fouxon1} I. Fouxon, B. Meerson,  M. Assaf, and E. Livne, Phys.
Rev. E \textbf{75}, 050301(R) (2007).
\bibitem{Fouxon2} I. Fouxon, B. Meerson,  M. Assaf, and E. Livne, Phys. Fluids \textbf{19}, 093303 (2007).
\bibitem{MFV} B. Meerson,  I. Fouxon, and A. Vilenkin, arXiv:0708.3085 [cond-mat.soft].
\bibitem{Kadanoff} L. P. Kadanoff, Physics Today \textbf{50}, 11 (1997).
\bibitem{BP} N.V. Brilliantov and T. P\"{o}schel, \textit{Kinetic Theory of Granular
Gases} (Oxford University Press, Oxford, 2004).

\bibitem{zeldovich} Ya. B. Zel'dovich and Yu. P. Raizer, \textit{Physics of Shock Waves and
High Temperature Hydrodynamic Phenomena}, vol. 1 (Academic Press,
New York, 1966).

\bibitem{Whitham} G. B. Whitham, \textit{Linear and Nonlinear Waves}
(Wiley, New York, 1974), Chapter 2.

\bibitem{Haff} P.K. Haff, J. Fluid Mech. \textbf{134}, 401 (1983).

\bibitem{Knudsencheck} The assumption of a small Knudsen number also breaks down close to the singularity.  However, the finite heat diffusion invalidates the solution much earlier, see Ref.~\cite{Fouxon2}.

\bibitem{poschel}  T. P\"oschel and T. Schwager, \textit{Computational Granular Dynamics} (Springer, Berlin, 2005).

\bibitem{blom} J.G. Blom and P.A. Zegeling, ACM Trans. Math. Software
\textbf{20}, 194 (1994).

\bibitem{Carnahan}  N. F. Carnahan and K. E. Starling, J. Chem. Phys. \textbf{51}, 635 (1969).

\bibitem{jenkins} J.T. Jenkins and M.W. Richman, Phys. Fluids \textbf{28}, 3485 (1985); Arch. Rat. Mech. Anal. \textbf{87}, 355
(1985).

\bibitem{Meerson}  B. Meerson, Rev. Mod. Phys. \textbf{68}, 215 (1996).

\end{thebibliography}
\end{document}